\documentclass[journal,twoside,web]{ieeecolor}
\usepackage{generic}
\usepackage{cite}
\usepackage{amsmath,amssymb,amsfonts}
\pdfoutput=1

\usepackage{threeparttable}
\usepackage{multirow}
\usepackage{tabularx}
\usepackage{makecell}
\usepackage{booktabs}

\usepackage{algorithm}
\usepackage{algorithmic}
\usepackage{graphicx}
\usepackage{textcomp}
\usepackage{enumerate}

\usepackage{pifont}
\usepackage{float}
\usepackage{stfloats}
\usepackage{balance}

\makeatletter
\let\NAT@parse\undefined
\makeatother
\usepackage{hyperref}  

\begin{document}
\title{ECG-CL: A Comprehensive Electrocardiogram Interpretation Method Based on Continual Learning} 
\author{Hongxiang Gao, Xingyao Wang, Zhenghua Chen, \IEEEmembership{Senior Member, IEEE}, Min Wu, \IEEEmembership{Senior Member, IEEE},  Jianqing Li and Chengyu Liu, \IEEEmembership{Senior Member, IEEE}
\thanks{This research was funded by the National Natural Science Foundation of China (62171123, 62211530112, 62201144 and 62071241), the National Key Research and Development Program of China (2022YFC2405600), the Postgraduate Research \& Practice Innovation Program of Jiangsu Province (KYCX20\_0088), the Fundamental Research Funds for the Central Universities (3201002106D), and the National Research Foundation, Singapore under its AI Singapore Programme (AISG2-RP-2021-027).
(Hongxiang Gao and Xingyao Wang contributed equally to this work.)
(Corresponding authors: Chengyu Liu and Zhenghua Chen (chengyu@seu.edu.cn, chen0832@e.ntu.edu.sg).) }
\thanks{Hongxiang Gao, Xingyao Wang, Jianqing Li and Chengyu Liu are with the State Key Laboratory of Digital Medical Engineering, School of Instrument Science and Engineering, Southeast University, Nanjing 210096, China (e-mails: \{hx\_gao, xingyao, ljq, chengyu\}@seu.edu.cn). }
\thanks{Hongxiang Gao, Zhenghua Chen, and Min Wu are with Institute for Infocomm Research, Xingyao Wang is with Institute of High Performance Computing, A*STAR, Singapore 138632, Singapore}}

\maketitle

\begin{abstract}

The value of Electrocardiogram (ECG) monitoring in early cardiovascular disease (CVD) detection is undeniable, especially with the aid of intelligent wearable devices. Despite this, the requirement for expert interpretation significantly limits public accessibility, underscoring the need for advanced diagnosis algorithms. Deep learning-based methods represent a leap beyond traditional rule-based algorithms, but they are not without challenges such as small databases, inefficient use of local and global ECG information, high memory requirements for deploying multiple models, and the absence of task-to-task knowledge transfer.
In response to these challenges, we propose a multi-resolution model adept at integrating local morphological characteristics and global rhythm patterns seamlessly. We also introduce an innovative ECG continual learning (ECG-CL) approach based on parameter isolation, designed to enhance data usage effectiveness and facilitate inter-task knowledge transfer.
Our experiments, conducted on four publicly available databases, provide evidence of our proposed continual learning method's ability to perform incremental learning across domains, classes, and tasks. The outcome showcases our method's capability in extracting pertinent morphological and rhythmic features from ECG segmentation, resulting in a substantial enhancement of classification accuracy.
This research not only confirms the potential for developing comprehensive ECG interpretation algorithms based on single-lead ECGs but also fosters progress in intelligent wearable applications. By leveraging advanced diagnosis algorithms, we aspire to increase the accessibility of ECG monitoring, thereby contributing to early CVD detection and ultimately improving healthcare outcomes.

\end{abstract}


\section{Introduction}
\label{sec:introduction}

\IEEEPARstart{C}{ardiovascular} diseases (CVDs) are the leading cause of global mortality and have a substantial impact on life satisfaction (WHO, 2019 \cite{who2019:CVDs}). Electrocardiogram (ECG) monitoring plays a critical role as a non-invasive method for the surveillance of cardiovascular risk events, enabling early detection of relevant ischemia and malignant arrhythmia.

\begin{figure}[t]
\centerline{\includegraphics[width=1\columnwidth]{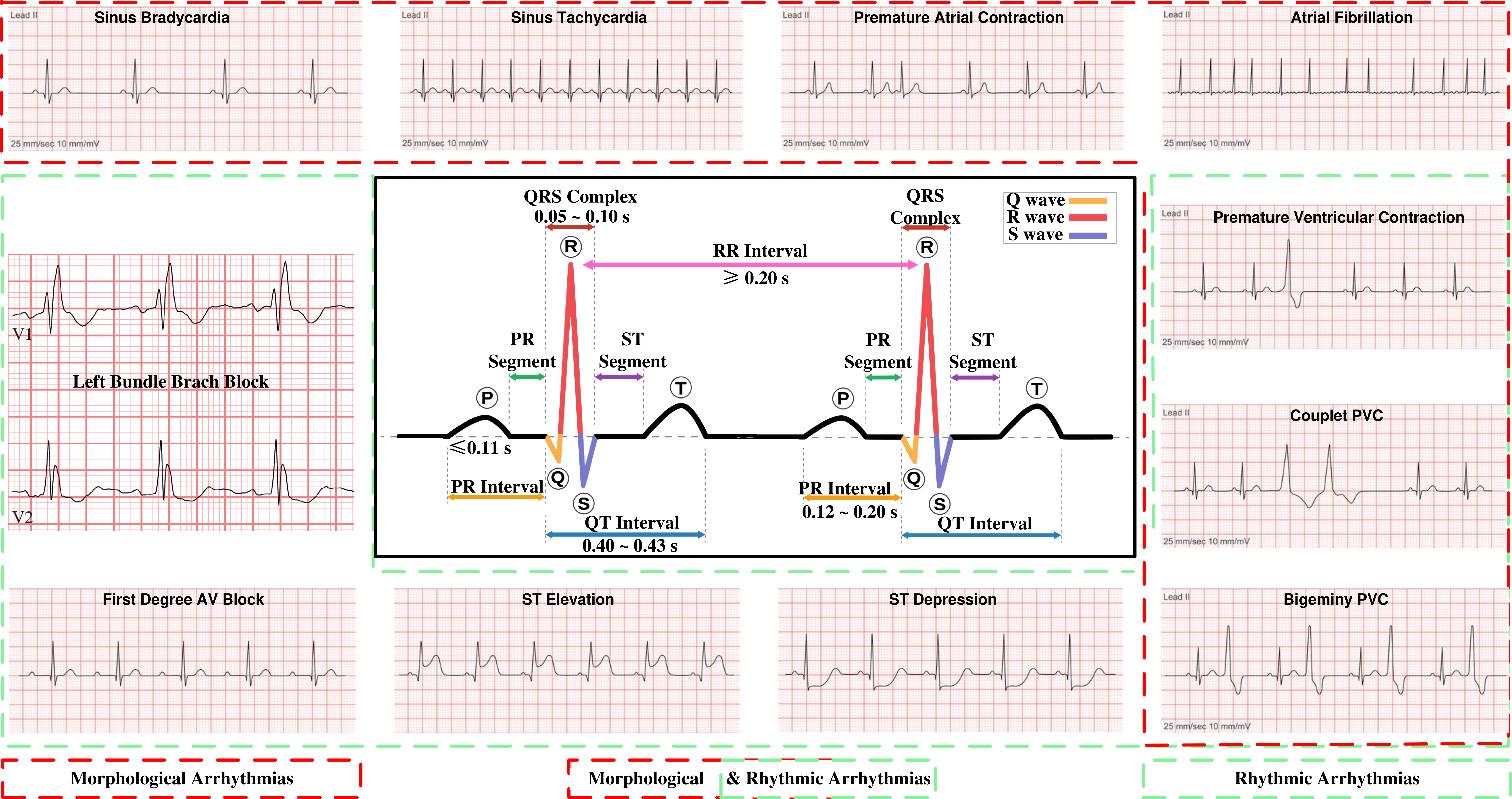}}
\caption{A schematic diagram detailing the analysis of an ECG. The central section illustrates the morphology of standard ECG waveforms and the standard interval information between two consecutive heartbeats. Circling this central depiction are examples of various cardiac arrhythmias. The arrhythmias encapsulated by the red dashed lines signify rhythm abnormalities, characterized by fluctuations in the RR intervals immediately preceding and following them. The irregularities enveloped by the green dashed lines denote morphological abnormalities, as indicated by variations in the QRS complex and other sub-waves. The intersection of the two boxes highlights instances wherein both rhythm and morphological abnormalities coexist.}
\label{fig1}
\end{figure}

The field of ECG interpretation has seen numerous transformative advancements over recent decades, parallel to the proliferation of open-source resources, such as databases \cite{gao2019open, liu2018:icbeb, Wagner2020:ptbxlphysionet, Gao2020} and algorithms \cite{wang2017time, pan1985real, cai2020qrs}. ECG analysis primarily involves the segmentation of key waveforms (QRS complexes, P waves, T waves) and the subsequent arrhythmia classification based on atypical morphological and rhythmic attributes. Prior research has primarily focused on QRS segmentation techniques leveraging the morphological, rhythmic, and amplitude characteristics of the QRS complex \cite{pan1985real}. These methods skillfully incorporate the morphological details of the QRS complex and the RR interval series into algorithms for arrhythmia classification, which are well-aligned with clinical knowledge.
Deep learning algorithms have significantly reduced the need to isolate QRS complexes prior to ECG event classification. These algorithms often use separate models for segmentation and classification tasks and have shown exceptional performance across diverse objectives \cite{ebrahimi2020review, yu2022ddcnn, sabor2022robust}. However, the widespread use of wearable devices faces three major challenges beyond the scope of these methods:
1. The lack of effective utilization of low- and high-resolution information for the smooth integration of contextual and semantic knowledge within the decision-making process.
2. The shortage of large ECG databases and inconsistencies in attribute types and data dimensions hinder the creation of universally applicable models.
3. The morphology and rhythm of the QRS waveform are essential for accurate QRS localization and are critical for accurate classification (as shown in Figure \ref{fig1}). Yet, an efficient mechanism for knowledge transfer in this context is notably lacking. Overcoming these challenges is essential to facilitate the universal adoption of wearable devices in ECG monitoring and interpretation.

The concept of learning at multiple resolutions has gained significant attention in the field of computer vision \cite{wang2020deep, chen2017deeplab}. Typically, convolutional neural networks (CNNs) for image classification follow a progression from high to low resolution, culminating in a final prediction. 
In the context of physiological signal analysis, specifically electrocardiograms (ECGs), the availability of large-scale databases \cite{Wagner2020:ptbxlphysionet, Wagner:2020PTBXL, liu2018:icbeb} and challenging QRS segmentation databases \cite{gao2019open, Gao2020} enabled the exploration of deep learning approaches that combine convolutional neural networks (CNNs) for feature extraction and recurrent neural networks (RNNs) for capturing temporal dependencies \cite{peimankar2021dens, liu2021deep}.
However, the aforementioned techniques primarily represent straightforward applications. ECG signals exhibit characteristic pseudo-periodicity and a distinct morphological distribution, which are crucial diagnosis criteria demanding further consideration. Cai \textit{et al.} \cite{cai2020qrs} proposed a solution that utilizes three convolutional branches with varying dilation rates to account for different visual scales, achieving the highest score in the 2019 China Physiological Signal Challenge (CPSC). To capture multi-resolution information, researchers have expanded the convolutional kernel and the dilation window \cite{wang2017time, Strodthoff:2020Deep}. However, these approaches independently capture distinct resolution characteristics without explicit interaction.

In this study, we introduce a novel CNN-based multi-resolution architecture that maintains high resolution throughout the learning process (Figure \ref{fig2} (a)). In each training phase, we retain the high-resolution branch and introduce a new low-resolution branch. By integrating the trained branches from the previous step, we ensure that both local and global characteristics are preserved in all branches. 
Additionally, we have developed a segmentation decoder (Figure \ref{fig2}(a.3)) based on multi-resolution and a classification decoder (Figure \ref{fig2}(a.4)) based on low-resolution.
Such a design is well-suited for handling pseudo-periodic signals that are controlled by the autonomic nervous system, such as respiratory signals, pulse wave signals, and heart sound signals.

In terms of data, the PhysioNet/Computing in Cardiology Challenge (CinC) 2020 \cite{alday2020classification} provides the largest collection of open-access ECG recordings from six medical centers. However, each database exhibits distinct cardiac abnormalities, and the class distribution follows a long-tail pattern, posing challenges for developing robust models. Currently, researchers utilize all available databases and train the final classifier using the entire range of categories. However, this approach exacerbates the performance by increasing the imbalance in categorical distribution.

Regarding the third challenge, several comparable works have been proposed. Salem \emph{et al.} \cite{Salem2018ECG} propose finetuning models by transferring information from computer vision tasks. Kuba \emph{et al.} \cite{weimann2021transfer} pretrain models on extensive ECG databases and optimize them for specific classification tasks. Raza \emph{et al.} \cite{raza2022designing} leverage federated knowledge to denoise ECG signals before transferring information for categorization. However, these approaches mainly focus on small databases and employ transfer learning in the fine-tuning mode. In contrast, Kiyasseh \emph{et al.} \cite{kiyasseh2021clinical} propose an ECG approach for continuous learning, considering data source acquisition, individual variation, and increasing categorization. To the best of our knowledge, no previous efforts have addressed the challenge of transferring information from segmentation to classification tasks.

In Figure \ref{fig3}, we present a solution to address the two challenges mentioned above using a parameter isolation-based continual learning approach. Firstly, we tackle the issues of data deficiency and label inconsistency by employing class-incremental continual learning, which integrates valuable information from each database.
This approach enables us to learn from multiple databases sequentially while preserving previously acquired knowledge. 
Secondly, to address the problem of knowledge transfer, we adopt a domain-incremental continual learning approach, where we sequentially learn single-lead, multi-lead segmentation, minority-class, and multiple-class categorization tasks. This process allows us to share generic features learned from previous tasks and generate specific features for the current task. Additionally, meaningful QRS complex segmentation can leverage information such as QRS duration, QRS morphology, RR interval, and other baseline features, which are crucial diagnoses references in clinical settings.

The main contributions of this paper are summarized as follows:
\begin{enumerate}
    \item We propose a novel CNN-based multi-resolution model designed to conserve both high-resolution morphological semantics and low-level rhythmical features in ECG interpretation. 
    The model incorporates a well-orchestrated convolution module, thereby effectively performing both classification and segmentation tasks.
    \item This paper pioneers the integration of segmentation and classification tasks within a unified framework. By leveraging generic features learned from previous tasks, the model enhances performance on downstream tasks. Moreover, the model cultivates task-specific features to address the current problem, thus providing a viable solution for databases with limited data availability.
    \item Extensive experiments are executed using four publicly accessible databases: CPSC 2019, 12-lead QRS, ICBEB 2018, and PTBXL. The results bear witness to the efficacy of the proposed ECG-CL approach in providing a comprehensive interpretation of ECG signals. Furthermore, the experiments corroborate the potential applicability of the approach in wearable devices, as reflected in the results of the single-lead experiment.
\end{enumerate}

\section{Related Works}

\subsection{Deep Learning-Based ECG Analysis}
The analysis of ECG data is a multi-faceted task, requiring techniques such as semantic segmentation \cite{he2020automatic, gabbouj2022robust} and the detection of rhythm and morphological anomalies \cite{hannun2019cardiologist, yu2022ddcnn}. With wearable ECG monitors becoming increasingly prevalent, deep learning methods have become the preferred approach over traditional rule-based techniques.
CNNs, RNNs, and combined CNN+RNN architectures are widely used in this context to capture the spatial attention and temporal dependencies inherent in ECG signals. Although ECG segmentation typically focuses on a single lead \cite{cai2020qrs}, classification often requires data from multiple leads. Despite this, existing deep learning methods treat these two tasks separately.
Several studies have leveraged heartbeat segmentation to improve accuracy and interpretability \cite{sabor2022robust, tison2019automated}. However, these approaches often rely on pre-existing heartbeat data or QRS localization algorithms. More importantly, they fail to fully integrate the tasks of ECG segmentation and classification, thus limiting their practicality.
Our research proposes a continual learning approach that addresses ECG analysis in a unified manner, merging the tasks of segmentation and classification to improve overall efficiency.

\subsection{Multi-resolution Neural Networks}
Multiresolution techniques have been widely explored in the field of computer vision \cite{zhao2017pyramid, wang2020deep, chen2017deeplab}. Some straightforward methods involve the parallel processing of multi-resolution data and the aggregation of output responses \cite{cai2020qrs}. Other techniques, such as U-Net \cite{ronneberger2015u}, use skip connections to progressively combine low-level and high-level features during the processes of downsampling and upsampling. Pyramid-based models leverage hierarchical features achieved through feature pooling. The HRNet \cite{wang2020deep} model retains high-resolution features throughout the training process. However, these models are primarily designed for image data, which are represented as square matrices.
In contrast, ECGs, which are typically recorded at a high sampling rate, present challenges in maintaining high-resolution data within the limited receptive fields of the network. Both local morphological information and global rhythm information are crucial for both segmentation and classification tasks in ECG analysis. To address these challenges, we propose a CNN-based multi-resolution network as an encoder with dedicated decoders for segmentation and classification tasks.

\subsection{Continual Learning}
Continual learning in machine learning is a dynamic approach that permits algorithms to learn and retain previously acquired knowledge while adjusting to new tasks, mimicking the learning process in humans \cite{GidoMvandeVen2019ThreeSF}. Various continual learning strategies have been proposed, including memory-based, regulation-based, and parameter isolation-based methods \cite{qu2021recent}. Each has its unique way of managing previously acquired knowledge and learning new tasks.

In the context of ECG interpretation, deep learning-based QRS segmentation and abnormality detection can benefit from the continual learning approach. Segmentation is generally treated as a binary classification problem, with continual learning over multiple databases helping in identifying new QRS morphologies. Classification tasks can be viewed as a class-incremental scenario, where various non-unified databases are integrated to form a comprehensive model for abnormality detection. Further, the transition from segmentation to classification can be seen as a domain-incremental scenario, leveraging segmentation features for classification tasks. Our work utilizes a parameter isolation-based continual learning strategy to address these scenarios in ECG analysis.

\section{Methods}

\begin{figure}[t]
    \centering
    \centerline{\includegraphics[width=0.9\columnwidth]{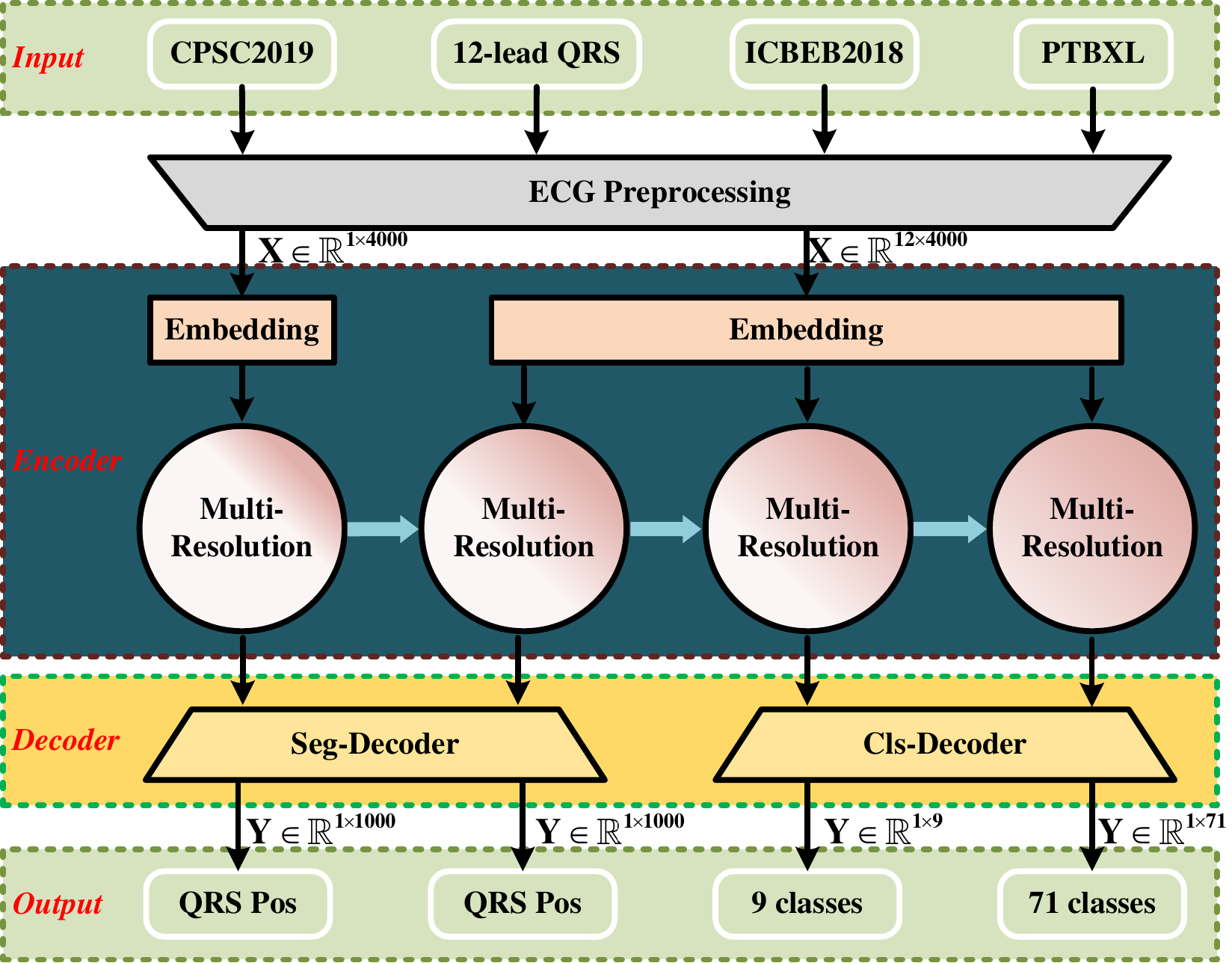}}
    \caption{ECG-CL Processing: An Overview of the Block Diagram.}
    \label{fig7}
\end{figure}

\begin{figure*}[t]
    \centering
    \centerline{\includegraphics[width=6.5in]{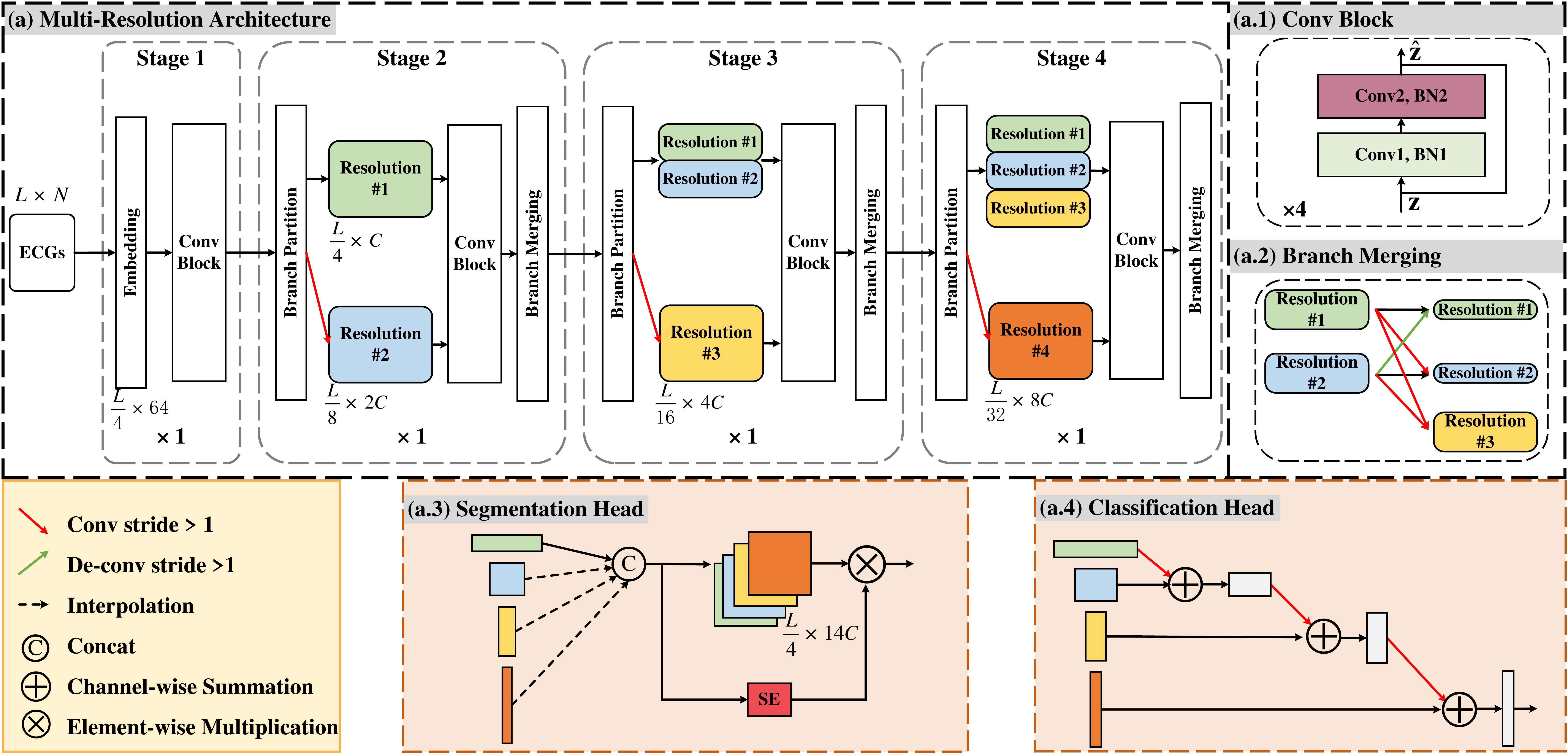}}
    \caption{The entire ECG processing backbone comprises the multi-resolution architecture (a) along with two decoders for segmentation (a.3) and classification (a.4). The multi-resolution architecture is divided into four stages, with each stage composed of modularized blocks. The branch partition module splits the current filters into two branches, with half retained for the current resolution and the other half allocated for a lower resolution. Following a convolution block (a two-layer residual module), the branch merging module (a.2) employs either strided convolution or deconvolution methods to integrate features of differing resolutions. ECG segmentation requires a high-resolution feature that interpolates low-resolution information. Conversely, ECG classification can be executed on low-resolution features by progressively transferring information from other resolutions to the lowest.}
    \label{fig2}
\end{figure*}

\subsection{Task Definition}
Our study introduces a universal multi-resolution architecture for ECG signal interpretation (ECG-MR), addressing ECG segmentation and classification, both crucial for wearable device functionality. Current ECG databases have limitations, such as small sizes and inconsistent labeling, and existing models cause storage and processing inefficiencies due to sequential deployment. We propose a domain and class-incremental continual learning approach to tackle these issues. Our ECG continual learning (ECG-CL) model leverages previously acquired knowledge to effectively adapt to new tasks. The overall learning process applied to the model is illustrated in Figure \ref{fig7} with the example of a comprehensive cross-domain continual learning task on four databases.

\subsection{Convolution-Based High-Resolution Modeling for ECGs}

\subsubsection{Multi-resolution Encoder} 
Figure \ref{fig2} (a) illustrates the Multi-Resolution architecture in a concise form, highlighting its key components. The architecture is designed for ECG interpretation tasks and operates on an input ECG signal ${\bf X} \in \mathbb{R}^{L \times N}$, where $L$ represents the signal length and $N$ denotes the number of leads.
The initial step involves a convolutional embedding layer that projects the raw ECG signal into a higher-dimensional space, reducing the length dimension by a factor of four. Notably, to accommodate ECGs obtained from non-standard devices and ensure compatibility across different input types, a $1 \times 1$ convolution layer is employed to project the ECGs to a standardized 12-channel representation.
Subsequently, the embedded features undergo a series of Conv blocks (depicted in Figure \ref{fig2} (a.1)), which preserve the first signal resolution ($C \times \frac{L}{4}$). This stage, along with the embedding module, is collectively referred to as "Stage 1".

To capture hierarchical representations, branch partition layers are introduced as the network deepens. The first branch partition layer maintains the initial embedded feature shape and applies stride convolutions to extract higher-level embeddings. This process reduces the feature dimensionality in each channel by half, resulting in a down-sampling of the resolution by a factor of two (i.e., $1 \times \frac{L}{8}$). The output dimension is set to twice the number of channels ($2C$). Subsequently, Conv blocks are applied for further feature transformation, with the resolution matching the input branch. A branch merging layer (as depicted in Figure \ref{fig2} (a.2)) is then employed to integrate local and global semantic information effectively. This set of operations, comprising branch partition, feature transformation, and branch merging, is denoted as "Stage 2".
This procedure is repeated twice more for "Stage 3" and "Stage 4", resulting in progressively lower resolutions of $1 \times \frac{L}{16}$ and $1 \times \frac{L}{32}$, respectively. Collectively, these stages generate a hierarchical representation that captures high-level semantic information while preserving fine-grained details at higher resolutions. Thus, the proposed Multi-Resolution architecture is well-suited for ECG interpretation tasks.

\subsubsection{Segmentation Decoder}
To address the tasks of ECG segmentation and classification, we propose segmentation and classification decoders that follow the main backbone of the network. ECG segmentation, which can be likened to an object detection task in time series, requires higher-resolution information for accurate decision-making. To achieve this, we interpolate the lower-resolution representations to match the shape of the highest-resolution representation and apply a Squeeze-and-Excitation (SE) module to enhance attention on effective semantics, as depicted in Figure \ref{fig2} (a.3). The segmentation decoder can be expressed as follows:
\begin{equation}
    {\bf Z} = \text{Concat}\Big({\bf z}_0, \mathcal{I}({\bf z}_1), \mathcal{I}({\bf z}_2), \mathcal{I}({\bf z}_3)\Big),
    \label{eq1}
\end{equation}
\begin{equation}
    \hat{\bf Z} = {\bf Z} \odot \text{SE}({\bf Z}),
\end{equation}
\begin{equation}
    {\bf O}_{\rm seg} = \text{Sigmoid}\Big(w \cdot (\text{AAP}(\hat{\bf Z})) + b\Big).
\end{equation}
Here, ${\bf z}_0 \in \mathbb{R}^{C \times \frac{L}{4}}, {\bf z}_1 \in \mathbb{R}^{2C \times \frac{L}{8}}, {\bf z}_3 \in \mathbb{R}^{4C \times \frac{L}{16}}$, and ${\bf z}_4 \in \mathbb{R}^{8C \times \frac{L}{32}}$. The concatenated feature representation is denoted as ${\bf Z} \in \mathbb{R}^{14C \times \frac{L}{4}}$, where $\mathcal{I}(\cdot)$ denotes the interpolation function, $\text{SE}$ denotes the SE module, $\odot$ denotes element-wise multiplication, $w$ and $b$ are the parameters of a fully connected layer, and $\text{AAP}(\cdot)$ denotes the Adaptive Average Pooling function. Finally, the output logits ${\bf O}_{\rm seg} \in \mathbb{R}^{\frac{L}{4}}$ with values greater than 0.5 indicate potential points of interest in the ECG segmentation task.

\subsubsection{Classification Decoder}
In a similar fashion, we design the classification decoder based on the multi-resolution output and transpose them to a low-dimensional representation for categorical decision, as illustrated in Figure \ref{fig2} (a.4). Instead of directly utilizing the lowest resolution representation as traditional classification models do, we progressively introduce the high-resolution representation into the lower resolution through strided convolution and channel-wise summation. The classification decoder can be expressed as follows:
\begin{equation}
    \hat{\bf z}_{i+1} = \textbf{SConv}({\bf z}_{i}) \oplus {\bf z}_{i+1}, \quad \text{s.t.} \quad i \in \{0, 1, 2\},
\end{equation}
\begin{equation}
    {\bf Z} = \hat{\bf z}_3,
\end{equation}
\begin{equation}
    {\bf O}_{\rm cls} = \text{Sigmoid}\Big(w \cdot \text{GAP}({\bf Z}) + b\Big).
\end{equation}
Here, $\textbf{SConv}$ denotes the strided convolution, $\oplus$ denotes channel-wise summation, $\text{GAP}(\cdot)$ denotes the global average pooling layer, and $w$ and $b$ represent the parameters of a fully connected layer. ${\bf O}_{\rm cls} \in \mathbb{R}^n$ denotes the logits for multi-label classification with $n$ classes.

\begin{figure}[t]
    \centering
    \centerline{\includegraphics[width=2.4in]{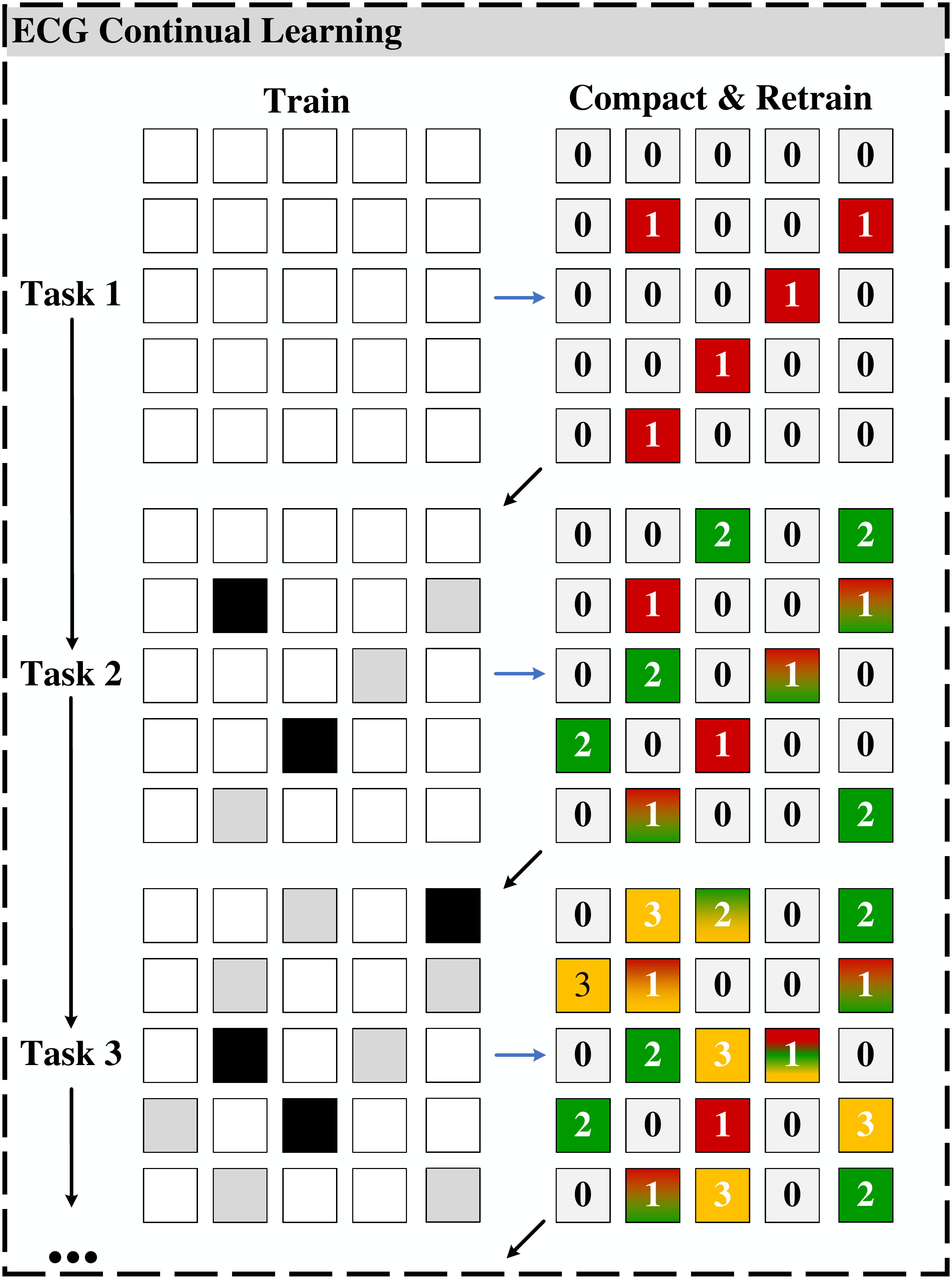}}
    \caption{Illustration of continual learning on model parameters. Empty cells in the training matrix are available for training, whereas grey cells are utilized primarily for forward training, and black cells are frozen. In the compact \& retrain matrix, $t$-filled cells indicate the weights for Task $t$, zero indicates released parameters, and the gradient color indicates the weights are utilized for more than one task.}
    \label{fig3}
\end{figure}

\subsection{ECG Continual Learning}
\subsubsection{Motivation} It is reasonable to assume the existence of generic concepts that characterize features across different domains (as depicted in Figure \ref{fig1}), which facilitates knowledge transfer between domains. We propose an iterative pruning method combined with the selection of generic weights.

This work adopts a parameter-isolated continual learning scenario, building upon the framework proposed in the PackNet paper \cite{ArunMallya2017PackNetAM}. The backbone model, along with the domain-specific decoders, is trained on a sequence of ECG interpretation tasks. To achieve a unified ECG interpretation framework, we employ a three-stage iterative process.
The multi-resolution architecture serves as the general backbone network for the feature encoder, making it shareable between segmentation and classification tasks. The proposed decoders, on the other hand, are task-specific and exclusive for dissimilar tasks, while being shareable within similar tasks.
The continual learning technique is applied to these shareable weights through an iterative process of train-compact-retrain for each task. This approach takes into account the complexity of general ECG interpretation, which involves data incremental learning for segmentation tasks, class-incremental learning for classification tasks, and domain incremental learning for transferring knowledge from segmentation to classification tasks.

\subsubsection{Continual Training}
Let $\mathcal{T}$ denote the set of all tasks, $\mathcal{T}^S$ denote the set of segmentation tasks, and $\mathcal{T}^C$ denote the set of classification tasks. Thus, we have $\mathcal{T} = \mathcal{T}^S \cup \mathcal{T}^C$. The indices $t$, $s$, and $c$ represent the current task index in the total set of tasks, in the segmentation tasks, and in the classification tasks, respectively. Therefore, we have $t = s + c$.

\textbf{Task} $\mathcal{T}_1$: In this case, illustrated in the first row of Figure \ref{fig3}, we train an initial network from scratch for the first task, denoted as $\mathcal{T}_1$. Subsequently, we prune the weights to a pre-defined sparsity level. To recover the performance that may have been affected by aggressive weight pruning, we apply a short fine-tuning step with a low learning rate. The preserved encoder weights for the first task are denoted as $W^P_{\text{MR}_1}$, while the released weights are denoted as $W^R_{\text{MR}_1}$. The same process is applied to the two decoders, where the preserved weights and released weights for the first segmentation task are $W^R_{\text{Seg}_1}$ and $W^P_{\text{Seg}_1}$, respectively. Similarly, for the first classification task, the preserved weights are denoted as $W^R_{\text{Cls}_1}$ and the released weights as $W^P_{\text{Cls}_1}$.

\textbf{Task} $\mathcal{T}_t \to \mathcal{T}_{t+1}$: Let's assume that we have completed the learning on tasks $\mathcal{T}_{1:t}$, which includes $s$ segmentation tasks and $c$ classification tasks. We denote the preserved model weights for segmentation tasks $\mathcal{T}^S_{1:s}$ as $W^P_s = W^P_{\text{MR}_{1:t}} \cup W^P_{\text{Seg}_{1:s}}$, and the preserved model weights for classification tasks $\mathcal{T}^C_{1:c}$ as $W^P_c = W^P_{\text{MR}_{1:t}} \cup W^P_{\text{Cls}_{1:c}}$. We then train weight-picking masks $M_s$ and $M_c$ on the preserved weights to select generic features from the previous tasks, facilitating the transfer of favorable knowledge to downstream tasks. 

When dealing with a new task $\mathcal{T}_{t+1}$, let us assume it is a classification task. We first train a network using the mask-picked weights $W^P_c \odot M_c$, which provide the generic knowledge, along with the released weights $W^R_c = W^R_{\text{MR}_t} \cup W^R_{\text{Cls}_c}$ to replenish the task-specific weights for $\mathcal{T}_{t+1}$. Notably, the gradient of the picked weights is set to zero to preserve the performance on past tasks. Subsequently, a prune and retrain step is added to create space for the next task and recover performance on the current task. This three-stage pattern is repeated until all tasks have been learned.

\begin{algorithm}[t]
    \renewcommand{\algorithmicrequire}{\textbf{Input:}}
	\renewcommand{\algorithmicensure}{\textbf{Output:}}
    \caption{Comprehensive ECG continual learning strategy.}
    \label{alg1}
    \begin{algorithmic}[1]
        \REQUIRE Pretrained encoder for task $\mathcal{T}_1$; segmentation decoder for task $\mathcal{T}^S_1$; classification decoder for task $\mathcal{T}^C_1$; training mode $m$ (\textit{Seg} or \textit{Cls}).
        \ENSURE The ECG interpreter for tasks $\mathcal{T}_1$ to $\mathcal{T}_T$.
        \WHILE{$t = 1$}
        \IF{$m_1$ == \textit{Seg}}
            \STATE Perform prune and retrain. Let the preserved weights be $W_1^P = W^P_{\text{MR}_1} \cup W^P_{\text{Seg}_1}$ and the released weights be $W_1^R = W^R_{\text{MR}_1} \cup  W^R_{\text{Seg}_1}$;
        \ELSIF{$m_1$ == \textit{Cls}}
            \STATE Perform prune and retrain. Let the preserved weights be $W_1^P = W^P_{\text{MR}_1} \cup W^P_{\text{Cls}_1}$ and the released weights be $W_1^R = W^R_{\text{MR}_1} \cup  W^R_{\text{Cls}_1}$;
        \ENDIF{}
        \ENDWHILE
        \IF{$t \leq T$}
        \STATE Assert $s$ segmentation tasks $c$ classifications tasks included, i.e., $t=s+c$;
        \IF{$m_t$ == \textit{Seg}}
            \STATE Preserved weights $W^P_{t-1} = W^P_{\text{MR}_{1:t-1}} \cup W^P_{\text{Seg}_{1:s-1}}$;
            \STATE Released weights $W^R_{t-1} = W^R_{\text{MR}_{1:t-1}} \cup W^R_{\text{Seg}_{1:s-1}}$.
        \ELSIF{$m_1$ == \textit{Cls}}
            \STATE Preserved weights $W^P_{t-1} = W^P_{\text{MR}_{1:t-1}} \cup W^P_{\text{Cls}_{1:c-1}}$;
            \STATE Released weights $W^R_{t-1} = W^R_{\text{MR}_{1:t-1}} \cup W^R_{\text{Cls}_{1:c-1}}$.
        \ENDIF
        \STATE Train task $\mathcal{T}_{t}$ using mask-picked weights and released weights: $({\bf M}_{t-1} \odot W^P_{t-1}) \cup {\bf W}_{t-1}^R$;
        \STATE Perform prune and retrain, update the preserved weights $W^P_t$ and released weights $W^R_t$ and the mask ${\bf M}_t$;
        \STATE ${t=t+1}$.
        \ENDIF
    \end{algorithmic}
\end{algorithm}

\subsubsection{Inference}
During inference, it is crucial to determine the task ID in order to retrieve task-specific parameters using preserved binary masks. In traditional multi-task inference, the loading of model weights often dominates the inference time, particularly in cascaded multi-task scenarios. However, our proposed continual learning technique only requires the model to be loaded once at the beginning, even in sequential multi-task settings. This sequential inference is achieved by leveraging task-ID-filled binary masks.

In the context of real-life ECG interpretation, let's consider a trained continual learning model with four tasks, namely "single-lead segmentation," "multi-lead segmentation," "limited-class classification," and "multi-class classification." An adaptive ECG interpretation system can be realized by incorporating both ECG segmentation and classification, depending on the shape of the input data. For a given ECG episode, the appropriate segmentation model for the corresponding lead can be selected, and classification can be performed using the most comprehensive classification model as the default choice. Furthermore, the task ID sequence can be manually defined to accommodate specific requirements and customize the ECG interpretation process accordingly.

The network train-prune-retrain procedures are performed iteratively for learning multiple new tasks. We summarize the overall learning process of ECG continual learning in Algorithm \ref{alg1}.

\section{Experimental Settings}
\label{Experiments}
In this section, we will provide a brief overview of the databases used for segmentation and classification and present the implementation details. Subsequently, we will evaluate the effectiveness of the proposed method by performing five types of continual learning schemes.

\subsection{Databases}
To evaluate the performance of the proposed architecture for ECG segmentation and classification, we select massive representative databases that are widely used in previous literature. Here is a summary of each database (also see in Table \ref{tbl1}):  \\
$\bigstar$ \textbf{ECG Segmentation Database} \\
\ding{182} \textbf{MIT-BIH Arrhythmia Database}

The MIT-BIH Arrhythmia Database (AR) \cite{moody2001impact} is the most widely recognized and extensively utilized database for QRS complex and arrhythmia beat detection. It consists of 48 records, each spanning 30 minutes, and sampled at a rate of 360 Hz. \\  \ding{183} \textbf{MIT-BIH Noise Stress Test Database}

The MIT-BIH noise stress test database (NSTD) \cite{moody1984noise} was constructed based on two clean recordings from the MIT-BIH-AD, intentionally contaminated with varying levels of noise. It encompasses six signal-to-noise ratios, resulting in a database containing 15 half-hour, two-lead ECG recordings sampled at 360 Hz. \\
\ding{184} \textbf{European ST-T Database}

The European ST-T database (EDB) \cite{taddei1992european} is an additional well-established resource commonly employed for QRS detection. It encompasses 90 beat-wise annotated recordings from 79 patients, obtained at a sampling rate of 250 Hz.  \\
\ding{185} \textbf{CPSC2019 QRS Database}

The CPSC2019 QRS database (CPSC2019) \cite{gao2019open} was released during the 2nd CPSC in 2019 and focuses on QRS detection in wearable dynamic ECGs. It consists of 5252 challenging single-lead ECGs (3232 for hidden test set) that contain various artifacts, impulse, strong noise, and abnormal morphology, is by far the most challenging database for QRS detection. Each recording is pre-processed to a length of 10 seconds with a sampling rate of 500 Hz. Three cardiologists meticulously annotated the QRS locations beat-by-beat to ensure the recognizability of the selected ECG clips.  \\
\ding{186} \textbf{12-lead QRS Database}

The 12-lead QRS database \cite{Gao2020} originating from the 1st CPSC in 2018, is specifically curated for QRS annotation research. It comprises 9364 beat-by-beat annotated electrocardiograms (ECGs) obtained by segmenting 12-lead ECG recordings into non-overlapping 10-second windows. The ECGs within the database are sampled at a rate of 500 Hz.  \\
$\bigstar$ \textbf{ECG Classification Database} \\
\ding{187} \textbf{ICBEB 2018 Database}

The ICBEB 2018 database \cite{liu2018:icbeb} released during CPSC 2018, comprises 6877 annotated 12-lead ECGs from six hospitals, encompassing eight arrhythmia types. Each recording for one patient lasts between 6 and 60 seconds and is annotated by up to three cardiologists, providing up to three statements per ECG. \\
\ding{188} \textbf{PTBXL Database}

The PTBXL database \cite{Wagner:2020PTBXL} is the largest publicly available ECG classification database to date. It consists of 21873 clinical 12-lead ECG records, each lasting for 10 seconds, collected from 18885 patients. The database follows the SCP-ECG standard ISO 91064 and includes 71 different statements categorized into 44 diagnoses, 12 rhythmic, and 19 morphological statements. Additionally, the database provides 5 super-diagnoses and 24 sub-diagnoses for diagnosis purposes. Each recording may have multiple statements associated with it.

\begin{table}
\centering
\renewcommand{\arraystretch}{1}
\begin{threeparttable}
    
\setlength\extrarowheight{1 pt}
\caption{Summary of the Used Databases for Both Segmentation (Seg) and Classification (Cls)}
\label{tbl1}
\setlength{\tabcolsep}{0.8mm}{
    \begin{tabular}{c|cccccc}
    \toprule[1.2pt]
                           & Database                & Fs (Hz) &  Length    &\#Leads & \#recordings  & \#labels \\
    \midrule
    \multirow{5}{*}{Seg}   & MIT-BIH-AR            &360    & 30 mins      & 2      & 48     & 119,000 \\ 
                           &   MIT-BIH-NSTD         &360    & 30 mins      & 2      & 15     & 26,370  \\
                           &   EDB                  &250    & 2 h          & 2      & 90     & 802,866 \\
                           &   CPSC2019             &500    & 10 s         & 1      & 5232   & 71,683  \\
                           &   12-lead QRS          &500    & 10 s         & 12     & 9364   & 122,996 \\
    \midrule
    \multirow{2}{*}{Cls}   & ICBEB 2018             & 500   & 6-60 s     & 12       & 6877   & 9\\
                           & PTBXL                  & 500   & 10 s       & 12       & 21873  & 71\\
    \bottomrule[1.2pt]
    \end{tabular}
    }
\end{threeparttable}
\end{table}

\subsection{Experimental Settings}

\subsubsection{Data preprocessing}

We immediately retrieve the 10-second ECG recordings or generate them using non-overlapping sliding windows to use in segmentation and classification training. As primary ECGs are effective in certain frequency bands, we do data pre-processing to convert raw ECG signals to the desired frequency domain (0.5 - 45 Hz). In addition, we normalize the 10-second signal such that the mean is zero and the variance is one. 

\subsubsection{Data splits}

For databases without pre-defined splits (e.g., MIT-BIH-AR \cite{moody2001impact}, NSTD\cite{moody1984noise}, European ST-T \cite{taddei1992european}, and 12-lead QRS databases \cite{Gao2020}), we have employed a 5-fold cross-validation strategy where folds are stratified by individual patients to avoid the leakage of patient-specific information.
For the CPSC2019 database \cite{gao2019open} which has a separate hidden test set defined during the challenge, we strictly followed this predefined split. Our model was trained on the provided training set and the final performance was reported based on the results from the hidden test set.
It is important to note that for all databases, we made sure the test set was only used for final evaluation to avoid overfitting or optimistic bias in the reported performance. The validation set, which was carved out from the training set, was used for hyperparameter tuning and model selection during the training phase.

For classification tasks, we followed the same data partitioning as in prior studies like \cite{Strodthoff:2020Deep} and \cite{Wagner2020:ptbxlphysionet}, ensuring fair and comparable experiments. The PTB-XL database's predefined splits were strictly adhered to as per the CinC 2020 guidelines.

\subsubsection{Implementation details}
Training for all models was conducted on two NVIDIA Tesla A100 GPUs, using PyTorch as the framework. We implemented a compact multi-resolution architecture consisting of four stages, each containing four Conv blocks, accommodating different resolution levels. To determine the optimal hyperparameters, we performed a search on the validation set, exploring the following ranges: (a) initial learning rate within the range [1e-4, 5e-1]; (b) batch size within range \{32, 64, 128, 256\} and hidden feature dimension within range \{4, 8, 12, 18\}.
We set the threshold for both segmentation and classification logits at 0.5. We employed a learning rate \textit{warm-up} strategy for the initial five epochs with a starting learning rate of 1e-6. The learning rate was halved every 30 epochs. 
During model pruning, we maintained a fixed learning rate of 0.0005. The pruning ratio was determined based on the number of tasks. We also evaluate the inference performance on a setup comprising of Intel I9-13900K CPU, Nvidia RTX 2080ti GPU, and 64GB RAM. Detailed hyperparameter settings and inference-related parameters can be found in Table \ref{tbl2}.

\begin{table}
\centering
\renewcommand{\arraystretch}{1}
\begin{threeparttable}
\setlength\extrarowheight{1 pt}
\caption{Hyperparameters Used for Model Training and Corresponding Model Inference Performance\tnote{*}}
\label{tbl2}
\setlength{\tabcolsep}{1mm}{
    \begin{tabular}{c|cccc|ccc}
    \toprule[1.2pt]
        Database       & BS        & Optim &  LR   & Epochs &\#Params (M)    &\#GFLOPS & SPS \\
    \midrule
        CPSC 2019     & 64         & Adam  & 0.001  &  50     & 0.79         & 0.34    & 210 \\
        12-lead QRS   & 128        & SGD   & 0.007  &  50     & 0.79         & 0.34    & 195\\
        ICBEB 2018    & 128        & SGD   & 0.015  &  50     & 1.63         & 0.36    & 165\\
        PTBXL         & 128        & SGD   & 0.500  &  200    & 1.69         & 0.36    & 149\\
    \bottomrule[1.2pt]
    \end{tabular}
    }
    \begin{tablenotes}
        \item[*] BS for Batch size, Optim for the optimizer, LR for learning rate, Fs for frequency of sampling, Params for the number of parameters, GFLOPS for Giga Floating-point Operations Per Second, SPS for samples per second.
    \end{tablenotes}
\end{threeparttable}
\end{table}

\subsubsection{Evaluation metrics}
To ensure that our results are comparable with previous works, we have adopted the True Positives (TP), False Positives (FP), False Negatives (FN), Sensitivity (SEN), and Positive Predictive (PP) and F1-score for assessing the performance of segmentation tasks.
In the case of multi-label classification tasks, we have reported the macro-averaged area under the receiver operating characteristic curve (macro-AUC)
The use of macro-averaged AUC as a performance metric aligns with the approach used in prior studies.

\section{Results and Discussion}
\subsection{Comparison with State-of-the-Art Approaches}
In this study, we conducted a comprehensive comparison of our proposed method against several state-of-the-art models for ECG segmentation and classification tasks across various databases. These include traditional rule-based algorithms \cite{pan1985real, sharma2019accurate, guendouzi2022qrs, tueche2021embedded, sabor2022robust, gungor2022stochastic, mourad2016efficient, nayak2019efficient}, which rely on inherent ECG signal characteristics, as well as deep learning-based methods \cite{cai2020qrs, karri2023real, wang2023causal, belkadi2021deep}. 
It is important to note that while rule-based algorithms can be applied to all databases, they often require specific parameter tuning for certain databases. Deep learning methods, on the other hand, require distinct training and testing databases, and thus their performance is evaluated on a subset of the data. 
The evaluation results in Table \ref{tbl3} revealed that while all methods can achieve high testing accuracy on traditional databases, their detection accuracy declined significantly on the NSTD and CPSC2019 databases due to the presence of uncertain noise and arrhythmia beats. 
Nevertheless, our method demonstrated superior or equivalent performance across all tasks. Particularly in the ECG classification task, our method achieved commendable performance as shown in Table \ref{tbl4}, highlighting the efficacy of our proposed model design.

Besides, results in Table \ref{tbl2} show that our model processes up to 201 segments per second for single-lead segmentation and achieves efficient rates even for complex tasks like 71-class classification with 12 leads at 145 segments per second. These results indicate that our model can fulfill the requirements for real-time diagnosis.
\begin{table*}[t]
    \centering
    \renewcommand{\arraystretch}{1}
    \begin{threeparttable}
    \setlength\extrarowheight{1 pt}
    \caption{Comparison of the State-of-the-Art ECG Segmentation Methods on Five Databases}
    \label{tbl3}
    \setlength{\tabcolsep}{4.3mm}{
    \begin{tabular}{cllllllll}
    \toprule[1.2pt]
    Database                                                                   & \multicolumn{1}{c}{Method} & \# R-peaks & \multicolumn{1}{c}{TP} & \multicolumn{1}{c}{FP} & \multicolumn{1}{c}{FN} & \multicolumn{1}{c}{SEN (\%)} & \multicolumn{1}{c}{PP(\%)} & \multicolumn{1}{c}{F1(\%)} \\ \hline
    \multirow{10}{*}{MIT-BIH-AR\cite{moody2001impact}}                                                 & P\&T (1985) \cite{pan1985real}              & 116137     & 115860                 & 507                    & 277                    & 99.67                        & 99.56                      & 99.61                      \\
                                                                              & Sharma (2022) \cite{sharma2022qrs}             & 110040     & 109910                 & 79                     & 130                    & 99.88                        & 99.93                      & 99.95                      \\
                                                                              & Guendouzi (2022) \cite{guendouzi2022qrs}           & 109494     & 109438                 & 95                     & 56                     & 99.95                        & 99.92                      & 99.93                      \\
                                                                              & Meghana (2023) \cite{karri2023real}            & 47579      & 47569                  & 5                      & 5                      & 100.00                       & 99.98                      & 99.99                      \\
                                                                              & Fabrice (2021) \cite{tueche2021embedded}            & 100427     & 100050                 & 298                    & 377                    & 99.65                        & 99.69                      & 99.67                      \\
                                                                              & Mohamed (2021) \cite{belkadi2021deep}             &            & 49707                  & -                      & -                      & 99.76                        & 99.24                      & 99.50                      \\
                                                                              & Nabil (2022) \cite{sabor2022robust}              & 109494     & 109342                 & 112                    & 152                    & 99.87                        & 99.77                      & 99.82                      \\
                                                                              & Cihan (2022) \cite{gungor2022stochastic}              & 109518     & -                      & -                      & -                      & 99.95                        & 99.96                      & 99.95                      \\
                                                                              & Cai (2020) \cite{cai2020qrs}                & -          & -                      & -                      & -                      & 99.94                        & 99.97                      & 99.95                      \\
                                                                              & ECG-MR                     & 109494     & 109406                 & 109                    & 88                     & 99.92                        & 99.90                      & 99.91                      \\ \midrule
    \multirow{4}{*}{NSTD\cite{moody1984noise}}                                                    & Sharma (2022) \cite{sharma2022qrs}             & 26010      & 25186                  & 1549                   & 655                    & 97.46                        & 94.20                      & 95.80                      \\
                                                                              & Cihan (2022) \cite{gungor2022stochastic}              & 25590      & -                      & -                      & -                      & 98.65                        & 99.11                      & 98.87                      \\
                                                                              & Cai (2020) \cite{cai2020qrs}                  & -          & -                      & -                      & -                      & 95.55                        & 92.93                      & 94.22                      \\
                                                                              & ECG-MR                     & 25590      & 24597                  & 1974                   & 93                     & 96.12                        & 92.57                      & 94.31                      \\ \midrule
    \multirow{6}{*}{\begin{tabular}[c]{@{}c@{}}European \\ ST-T  \cite{taddei1992european}\end{tabular}} & Mourad (2016) 
    \cite{mourad2016efficient}             & 788772     & 786012                 & 1258                   & 2760                   & 99.65                        & 99.84                      & 99.74                      \\
                                                                              & Nayak (2019) \cite{nayak2019efficient}               & 790560     & 789536                 & 1117                   & 1024                   & 99.87                        & 99.86                      & 99.86                      \\
                                                                              & Cihan (2022) \cite{gungor2022stochastic}              & 790558     & -                      & -                      & -                      & 99.93                        & 99.97                      & 99.95                      \\
                                                                              & Mohamed (2021) \cite{belkadi2021deep}            & 679795     & 677756                 & 13479                  & 2039                   & 99.70                        & 98.05                      & 98.87                      \\
                                                                              & Yun (2022) \cite{yun2022robust}  & 790560  & - & - & - & 99.85 & 99.91 & 99.88  \\
                                                                              & ECG-MR                    & 790558          & 789910                      & 950                      & 633                  & 99.92                        & 99.88                      & 99.90                      \\ \midrule
    \multicolumn{1}{l}{\multirow{7}{*}{CPSC2019\cite{gao2019open}}}                             & P\&T (1985) \cite{pan1985real}                       & 44274      & 41586                  & 2851                   & 3389                   & 92.46                        & 93.58                      & 93.02                      \\
    \multicolumn{1}{l}{}                                                      & Cai (2020) \cite{cai2020qrs}                  & 19455      & -                      & -                      & -                      & 99.70                        & 99.83                      & 99.76                      \\
    \multicolumn{1}{l}{}                  & {He (2020) \cite{he2020automatic}}    & {29437}                     & {28003}  & {6140}  & {1434}  & {95.13}  & {82.03}   & {88.09}  \\
    \multicolumn{1}{l}{}                  & {Ivora (2022) \cite{ivora2022qrs}}    & -                     & -  & -  & -  & -  & -   &{ 94.40}   \\
    \multicolumn{1}{l}{}                                                      & Wang (2023) \cite{wang2023causal}                & 44274      & -                      & -                      & -                      & 99.26                        & 99.45                      & 99.35                      \\
    \multicolumn{1}{l}{}                                                      & ECG-MR                     & 44274      & 43920                  & 477                    & 354                    & 99.20                        & 98.14                      & 99.04                      \\
    \multicolumn{1}{l}{}                                                      & ECG-CL                     & 44274      & 43943                  & 441                    & 331                    & 99.25                        & 99.01                      & 99.12                      \\ \hline
    \multicolumn{1}{l}{\multirow{5}{*}{12lead\_QRS\cite{Gao2020}}}                          & P\&T (1985) \cite{pan1985real}                       & 114182     & 109658                 & 7161                   & 4524                   & 96.04                        & 93.87                      & 94.94                      \\
    \multicolumn{1}{l}{}                                                      & {Cai (2020) \cite{cai2020qrs}}                  &{ 114182}      & {111953}                      & {661}                      & {718}                      & {99.37}                        & {99.41}                      & {99.39}                      \\
    \multicolumn{1}{l}{}                                                      & {Wang (2023)} \cite{wang2023causal}                & {114182}      & {112199}                      & {792}                      & {472}                      & {99.58}                        & {99.30}                      & {99.44}                      \\
    \multicolumn{1}{l}{}                                                      & ECG-MR                     & 10541      & 10531                  & 13                     & 10                     & 99.90                        & 99.93                      & 99.91                      \\
    \multicolumn{1}{l}{}                                                      & ECG-CL                     & 10541      & 10538                  & 2                      & 3                      & 99.98                        & 99.98                      & 99.98                      \\ \bottomrule[1.2pt]
    \end{tabular}}
    \end{threeparttable}
    \end{table*}

    \begin{table}[t]
    \centering
    \renewcommand{\arraystretch}{1}
    \begin{threeparttable}        
    \setlength\extrarowheight{1 pt}
    \caption{Comparison with the State-of-the-Art ECG Classification Methods on Two Databases}
    \label{tbl4}
    \setlength{\tabcolsep}{5mm}{
    \begin{tabular}{cll}
    \toprule[1.2pt]
    Database                    & \multicolumn{1}{c}{Method} & \multicolumn{1}{c}{AUC (\%)} \\ \midrule
    \multirow{8}{*}{ICBEB2018\cite{liu2018:icbeb}} & lstm(2020) \cite{Wagner2020:ptbxlphysionet}                 & 96.40                        \\
                               & xresnet1d101 (2020) \cite{Wagner2020:ptbxlphysionet}         & 97.40                        \\
                               & resnet1d\_wang (2020) \cite{Wagner2020:ptbxlphysionet}       & 96.90                         \\
                               & ensemble (2020) \cite{Wagner2020:ptbxlphysionet}             & 97.50                         \\
                               & Yang (2023) \cite{yang2023multi}                 & 95.80                        \\
                               & {ASTCL (2023) \cite{wang2023adversarial}}     & {84.24} \\
                               & ECG-MR                     & 97.30                        \\
                               & ECG-CL                     & 98.03                        \\ \midrule
    \multirow{11}{*}{PTBXL\cite{Wagner2020:ptbxlphysionet}}     & lstm (2020) \cite{Wagner2020:ptbxlphysionet}                 & 90.70                         \\
                               & xresnet1d101 (2020) \cite{Wagner2020:ptbxlphysionet}         & 92.50                         \\
                               & resnet1d\_wang (2020) \cite{Wagner2020:ptbxlphysionet}       & 91.90                         \\
                               & ensemble (2020) \cite{Wagner2020:ptbxlphysionet}             & 92.90                         \\
                               & Zhang (2021) \cite{zhang2021mlbf}               & 93.40                        \\
                               & Yang (2023) \cite{yang2023multi}                & 92.89                        \\
                               & Eedara (2022) \cite{prabhakararao2021multi}              & 87.11                        \\
                               & {CRT (2023) \cite{zhang2023self}}    & {89.22} \\
                               & {ASTCL (2023) \cite{wang2023adversarial}}   & {82.03} \\
                               & ECG-MR                     & 92.18                        \\
                               & ECG-CL                     & 92.73                        \\ \bottomrule[1.2pt]
    \end{tabular}}
    \end{threeparttable}
    \end{table}

\subsection{Continual Learning Performance}
We have conducted four types of continual learning experiments, tailoring the scenarios derived from the databases to mimic realistic applications. For all experiments, model performance for each task is evaluated once all tasks have been learned. We have chosen to utilize ResNet\_wang \cite{wang2017time} to replace MR architecture for both segmentation and classification, given its exceptional performance on ECG abnormality classification tasks \cite{Wagner:2020PTBXL} and its efficient use of model parameters. 
Continual learning presents two main advantages: the prevention of catastrophic forgetting and the promotion of forward knowledge transfer. To validate the first advantage, we carried out a "finetune" experiment. Here, we sequentially finetuned the previously well-trained model on the current database and reported the results for each task using the final model weights. It is important to note that like continual learning, we only fine-tuned the parameters that were shareable across tasks.
To evaluate the ability for forward knowledge transfer, we performed a "scratch" experiment. In this setting, each task was trained with an independent model from scratch, and the performance was evaluated individually.

\subsubsection{Comprehensive cross-domain continual learning}
Firstly, we set out to develop the most comprehensive ECG interpretation method by implementing continual learning on a sequence of four tasks. Guided by the principle of training the model from low-level to high-level tasks and transitioning from simple to complex tasks, we arranged the \textit{task} order and corresponding databases as follows:
\begin{enumerate}[{\bf 1.}]
    \setlength{\itemindent}{1.1em}
    \item \textit{single-lead segmentation}: CPSC 2019;
    \item \textit{multi-lead segmentation}: 12-lead QRS;
    \item \textit{fewer class classification}: ICBEB 2018;
    \item \textit{multiple class classification}: PTBXL-all.
\end{enumerate}

\begin{table}[t]
    \centering
    \renewcommand{\arraystretch}{1}
    \begin{threeparttable}
    \setlength\extrarowheight{1 pt}
    \caption{Overall Discriminative Performance of ECG Continual Learning Algorithms on Sequential Tasks\tnote{*} in Terms of F1\_score for Segmentation Tasks and Macro AUC for Classification Tasks. For Each Task, the Best-Performing Value is Marked in Bold Face. CL is an Abbreviation of Continual Learning.}
    \label{tbl5}
    \setlength{\tabcolsep}{0.8mm}{
        \begin{tabular}{cccccc}
            \toprule[1.2pt]
             Model                                      & Mode        & CPSC2019  & 12-lead QRS   & ICBEB 2018  & PTBXL \\
            \midrule 
            \multirow{3}{*}{\makecell[c]{ResNet\_wang}} & Scratch     & 98.83      & 99.93         & 96.90       & 91.90     \\
                                                        & Finetune    & 91.24      & 97.39         & 92.30       & 92.11     \\
                                                        & CL          & 98.87      & \bf 99.98     & 96.93   & 92.56     \\
            \midrule
            \multirow{3}{*}{\makecell[c]{ECG-CL}}         
                                                        & Scratch     & 99.04      & 99.91        & 97.30       & 92.18     \\
                                                        & Finetune    & 91.22      & 96.99        & 92.87       & 92.35     \\
                                                        & CL          &\bf 99.12   &\bf 99.98     & \bf 98.03   & \bf 92.73     \\
                                                        
        \bottomrule[1.2pt]
        \end{tabular}}
        \begin{tablenotes}
            \item [*] Task order: CPSC2019 (\textit{single-lead segmentation}) $\to$
    12-lead QRS (\textit{multi-lead segmentation}) $\to$ ICBEB2018 (\textit{fewer class classification}) $\to$ PTBXL (\textit{multiple class classification}) 
        \end{tablenotes}
    \end{threeparttable}
\end{table}

\begin{figure}[t]
    \centering
    \centerline{\includegraphics[width=0.8\columnwidth]{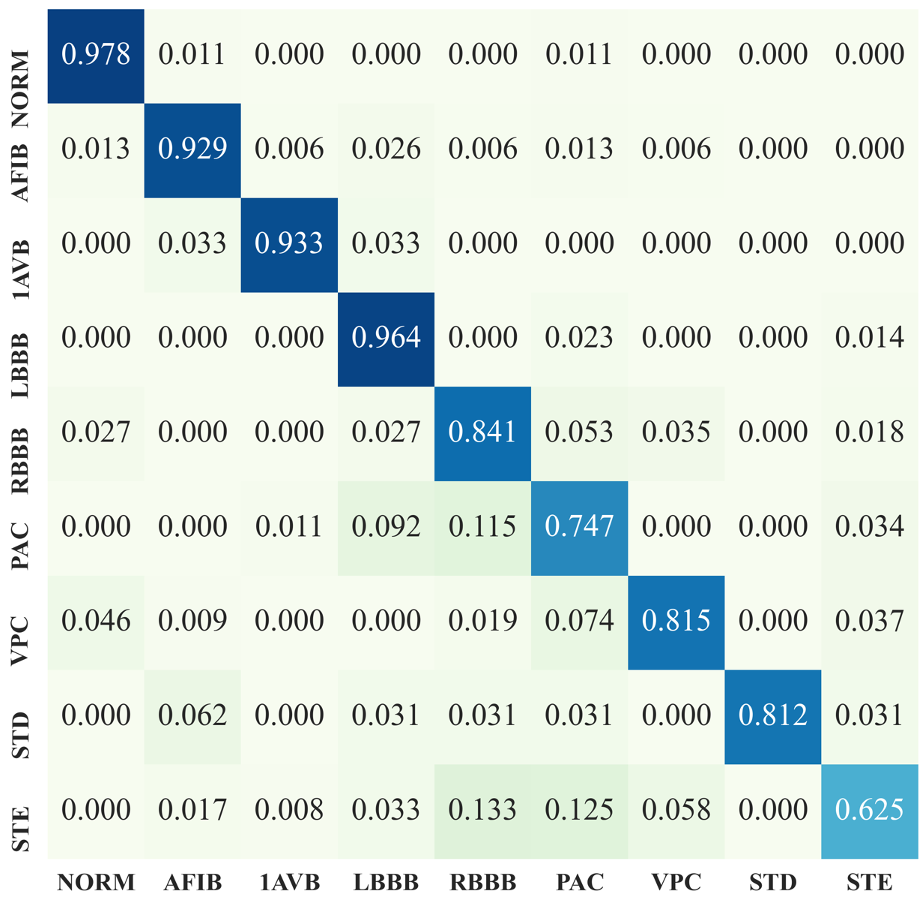}}
    \caption{Confusion Matrix Analysis for Nine-Class Classification on ICBEB Database.}
    \label{fig4}
\end{figure}

\begin{figure*}[t]
    \centering
    \centerline{\includegraphics[width=1.8\columnwidth]{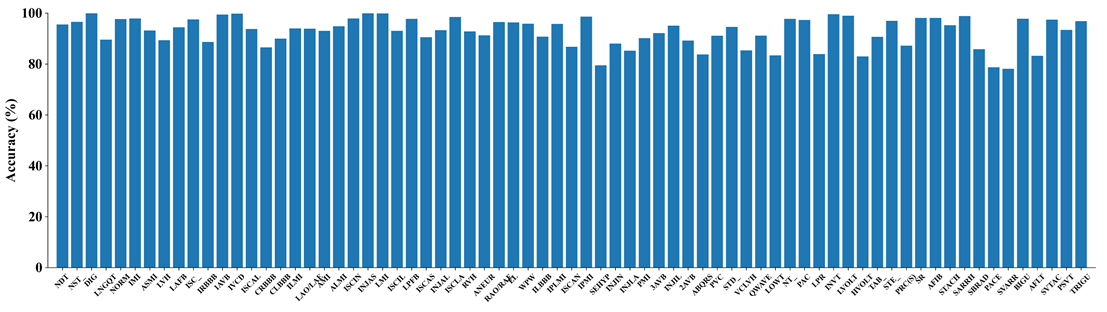}}
    \caption{Class-specific Macro-AUC Analysis on PTBXL Database.}
    \label{fig5}
\end{figure*}

Our experiments arranged the databases in a sequential manner, adhering to the pre-set order. In this arrangement, each downstream task generates task-specific features while simultaneously leveraging the general features learned from previous tasks. As a result, we anticipate the performance to surpass that of models trained from scratch. Interestingly, the accuracy on the CPSC 2019 database after pruning slightly exceeds that of the model trained from scratch, which may be attributed to the weight regularization capabilities of pruning. The confusion matrix on ICBEB2018 and class-wise AUCs are also provided in Figure \ref{fig4} and Figure \ref{fig5}.

Table \ref{tbl5} illustrates the experimental results of our ECG-CL method alongside the multi-resolution architecture. The results indicate that the continual learning method outperforms models trained from scratch in downstream tasks, reinforcing our assertion that our model can effectively transfer general knowledge from preceding tasks. It is worth noting that the storage requirements for training multiple models from scratch are several times greater than those for a single continual learning model. While fine-tuning on new tasks yields comparable results for the final two classification tasks, it fails to preserve distinct knowledge from the initial two tasks, thus leading to catastrophic forgetting.

\subsubsection{Continual Learning on Three Major Categories}

From a clinical perspective, ECG statements are typically interpreted based on the P-QRS-T morphology and wave duration information. The PTBXL database is categorized into three non-mutually exclusive types: diag (short for diagnoses statements such as "anterior myocardial infarction", 44 classes), form (related to notable changes in specific segments within the ECG such as "abnormal QRS complex", 12 classes), and rhythm (related to particular changes in rhythm such as "atrial fibrillation", 19 classes).
Confining the task to the classification of ECG abnormalities, PTBXL stands as the current largest database of abnormalities. It is reasonable to presume that any existing ECG abnormalities not listed, or to be identified in the future, could fall under one of these three categories. Thus, we trained the continual learning model on the form, rhythm, and diagnoses categories of the PTBXL database. Below, we describe the three types of \textit{tasks} and their corresponding databases.

\begin{enumerate}[{\bf 1.}]
    \setlength{\itemindent}{1.1em}
    \item \textit{12-lead form}: PTBXL-form;
    \item \textit{12-lead rhythm}: PTBXL-rhythm;
    \item \textit{12-lead diagnoses}: PTBXL-diag.
\end{enumerate}

\begin{table}[t]
    \centering
    \renewcommand{\arraystretch}{1}
    \begin{threeparttable}
    \setlength\extrarowheight{1 pt}
    \caption{Classification Performance of the Continual Learning Model on Three Descriptive Anomaly Tasks on PTBXL in Terms of Macro AUC.}
    \label{tbl6}
    \setlength{\tabcolsep}{1.4 mm}{
    \begin{tabular}{cccccc}
        \toprule[1.2pt]
         Model                                      & Mode        & PTBXL-form  & PTBXL-rythm & PTBXL-diag \\
        \midrule
        \multirow{2}{*}{\makecell[c]{ResNet\_wang}} & Scratch     & 87.58      & 94.60         & 93.60        \\
                                                    & CL          & 88.00      & 95.97         & \bf 94.27       \\
        \midrule
        \multirow{2}{*}{\makecell[c]{ECG-CL}}       & Scratch     & 87.59      & \bf 96.29     & 92.29     \\
                                                    & CL          & \bf 88.36  & \bf 96.29     & 93.13      \\
        \bottomrule[1.2pt]
    \end{tabular}}
    \end{threeparttable}
\end{table}

\begin{figure}[t]
    \centering
    \centerline{\includegraphics[width=0.8\columnwidth]{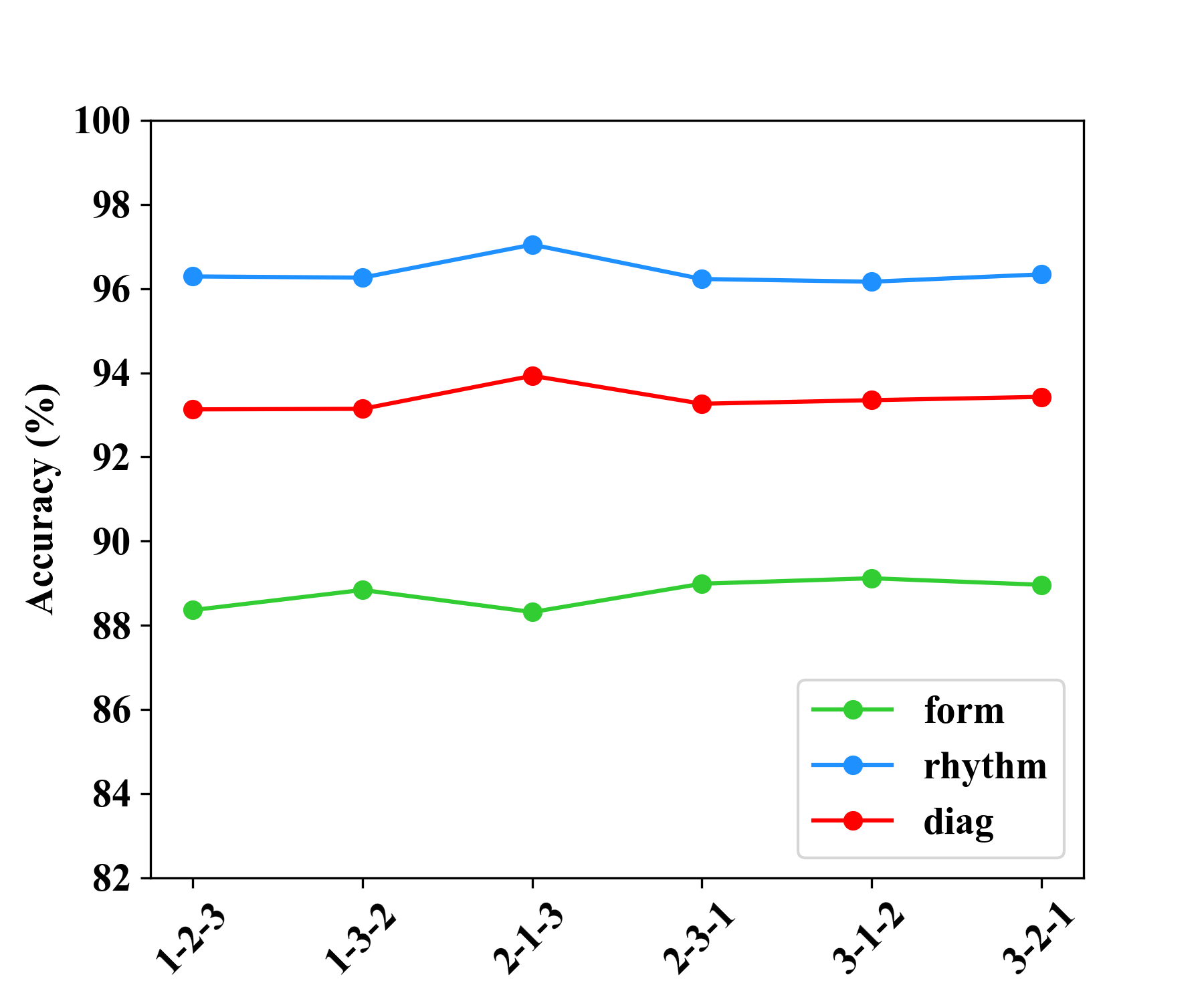}}
    \caption{Performance Evaluation of Continual Learning with Different Learning Sequence Orders on Three Descriptive Anomaly Tasks on PTBXL in Terms of macro-AUC.}
    \label{fig6}
\end{figure}

\begin{table}[t]
    \centering
    \begin{threeparttable}
    \setlength\extrarowheight{1 pt}
    \caption{Classification Performance of the Continual Learning Model on Five Diagnoses Tasks on PTBXL in Terms of Macro AUC.}
    \label{tbl7}
    \setlength{\tabcolsep}{1.8mm}{
    \begin{tabular}{ccccccc}
        \toprule[1.2pt]
         Model                                      & Mode        & NORM      & STTC           & HYP         & CD        & MI  \\
        \midrule
        \multirow{4}{*}{\makecell[c]{ResNet\_wang}} & Scratch     & 94.97     &  91.53         & 90.48        & 92.73     & 93.58      \\
                                                    & GEM         &  95.12   & \bf 92.34    & 90.77    & 92.96   & 93.58  \\
                                                    & ER-MIR      &  95.38  & 92.11 & 91.92 & 93.08 & 93.68  \\
                                                    & CL          & 95.20     & 86.70         & 92.63        & 93.33     & 93.40      \\
        \midrule
        \multirow{2}{*}{\makecell[c]{ECG-CL}}       & Scratch     & 96.25      & 88.25         & 94.44       & 94.12     & 93.67\\
                                                    & CL          & \bf 96.77      & 90.18         & \bf 95.55       & \bf 95.99     & \bf 93.81\\
        \bottomrule[1.2pt]
    \end{tabular}}
    \end{threeparttable}
\end{table}

\begin{table}[t]
    \centering
    \begin{threeparttable}
    \setlength\extrarowheight{1 pt}
    \caption{Performance of the Comprehensive ECG Continual Learning Model for Application of Wearable Smart Devices on Four Single-Lead databases in Terms of F1\_score for Segmentation Task and Macro AUC for Classification Task.}
    \label{tbl8}
    \setlength{\tabcolsep}{1.2mm}{
    \begin{tabular}{cccccc}
        \toprule[1.2pt]
         Model                                      & Mode  & CPSC 2019      & rhythm & QRS form  & ST-T form \\
        \midrule 
        \multirow{2}{*}{\makecell[c]{ResNet\_wang}} & Scratch     & 98.83    & 99.77    &  98.23           & 99.65     \\
                                                    & CL          & 98.87    & 99.70    &  98.66           & \bf 99.80     \\
        \midrule
        \multirow{2}{*}{\makecell[c]{ECG-CL}}       & Scratch     & 99.04    &  99.72     & 98.36          & 99.69      \\
                                                    & CL          & \bf 99.12& \bf 99.81  & \bf 98.75      & 99.76      \\
        \bottomrule[1.2pt]
    \end{tabular}}
    \end{threeparttable}
\end{table}

Given that there is no clear hierarchical relationship between these three tasks, we present their performance results in Table \ref{tbl6}, following the sequence mentioned above. Additionally, we explore the performance variations across six different curriculum learning schemes in Figure \ref{fig3}. As depicted in Figure \ref{fig6}, the order of tasks can slightly impact overall performance. This effect may be attributed to the similarity of classification tasks.

\subsubsection{Class-incremental Learning From Exclusive Categories}

Multi-lead ECG signals used in clinical applications must ultimately contribute to reliable diagnoses that can assist healthcare providers in improving efficiency or transforming the use of multi-lead wearables like the Holter monitor. The PTBXL database encompasses 44 commonly derived clinical diagnoses, which are subdivided into five super-diagnosis. However, there are still many diagnosis details not included, such as malignant ventricular fibrillation, electrical axis deviation, etc. 
To accommodate all future potential new super-diagnoses or sub-diagnoses, we design a task-incremental continual learning scheme. The five super-diagnosis provided in PTBXL are directly utilized here to evaluate the model's continual learning capability. 
Generally, ST-T changes are waveform phenomena that could be associated with conduction block, myocardial infarction, and hypertrophy. Pathological ST-T changes are not linked with other types. Myocardial infarction could be a trigger for conduction block and hypertrophy, while hypertrophy could cause some conduction disturbances. Based on this knowledge, we set the sequential order as follows:
\begin{enumerate}[{\bf 1.}]
    \setlength{\itemindent}{1.1em}
    \item \textit{normal (NORM)}: PTBXL-NORM;
    \item \textit{ST-T change (STTC)}: PTBXL-STTC;
    \item \textit{hypertrophy (HYP)}: PTBXL-HYP;
    \item \textit{conduction disturbance (CD)}: PTBXL-CD;
    \item \textit{myocardial infarction (MI)}: PTBXL-MI.
\end{enumerate}
For each super-diagnosis, we use the sub-diagnoses (23 in total) as training samples. Table \ref{tbl7} presents the comparison between the two architectures in terms of the three training modes. The overall trend aligns with previous experiments, and the results from the continual learning process significantly outperform those from training from scratch. This can be attributed to the fact that basic cardiac abnormalities often occur in tandem with other pathological changes.

Besides, we have indeed conducted comparisons with the Gradient Episodic Memory (GEM) method and the Memory Informed Rehearsal (MIR) method. We adopted the ResNet1d-Wang as the backbone architecture for all three methods and conducted our experiments on five super-diagnosis databases. We set the buffer memory for GEM and MIR at 256, and it is important to note that we only compared our method with the Experience Replay (ER) variant of the MIR method, which was found to have superior performance in the aforementioned study. As illustrated in Table \ref{tbl7}, GEM lags slightly behind the other two methods as it only stores previously trained samples in the buffer memory, thereby constraining its performance by the size of the memory. This limitation also affects MIR, although its performance is somewhat boosted by its strategy of selecting challenging samples from previous tasks. Nevertheless, both methods must retain a portion of learned samples, which introduces a significant storage burden, particularly for larger tasks. In summary, our parameter-isolated continual learning method exhibits superior performance, demonstrating the suitability of continual learning strategies for comprehensive ECG learning tasks aimed at solving the current challenges in this field.

\subsubsection{Continual Learning for Wearable Application}

Wearable devices like smartwatches and portable ECG monitors have gained popularity for monitoring heart health and detecting conditions like atrial fibrillation.
Single-lead ECGs can diagnose common arrhythmias such as premature atrial contraction (PAC), premature ventricular contraction (PVC), atrial fibrillation (AF), and ST-T changes. In this study, we performed continual learning from single-lead segmentation to classify rhythmic and morphological abnormalities. Rhythmic classification, based on heartbeat segmentation, is simpler compared to morphological classification, which involves QRS complexes and ST-T abnormalities. We used the ICBEB2018 database for abnormal rhythm classification, including PAC, PVC, and AF. For QRS morphological abnormalities, we utilized PVC, right bundle branch block (RBBB), and left bundle branch block (LBBB), while ST elevation (STE) and ST depression constituted the ST-T change database. The lead-I signal, which represents the electrical potential difference between the left and right hand, was adopted, noting that bundle branch block (BBB) exhibits significant abnormalities in this lead. The sequential order is as follows:
\begin{enumerate}[{\bf 1.}]
    \setlength{\itemindent}{1.1em}
    \item \textit{1-lead segmentation}: CPSC 2019;
    \item \textit{1-lead rhythmic classification}: ICBEB2018-rhythm;
    \item \textit{1-lead QRS form classification}: ICBEB2018--form;
    \item \textit{1-lead ST-T form classification}: ICBEB2018--STTC.
\end{enumerate}
Table \ref{tbl8} presents the comprehensive results of applying continual learning to single-lead data for prospective wearable device applications. The results confirm the feasibility of integrating a sequence of tasks into a single model for use in wearable devices.

\begin{table}[t]
    \renewcommand{\arraystretch}{1}
    \begin{threeparttable}
    \setlength\extrarowheight{1 pt}
    \caption{ECG-CL Ablation Study: F1-score for Segmentation Tasks and Macro-AUC for Classification Tasks across Four ECG Databases.}
    \label{tbl9}
    \setlength{\tabcolsep}{1.6mm}{
    \begin{tabular}{ccccc}
    \toprule[1.2pt]
    \multicolumn{1}{c}{Method} & CPSC2019 & 12lead-QRS & ICBEB2018 & PTBXL \\ \midrule
    ECG-CL                     & 99.12    & 99.98      & 98.03     & 92.73 \\
    w/o CL                     & 99.04    & 99.91      & 97.30     & 92.18 \\
    w/o MR                     & 98.87    & 99.98      & 96.93     & 92.56 \\
    w/o CL \& MR               & 98.96    & 99.82      & 96.90     & 91.90 \\
    w/o Decoder \& CL          & 99.01    & 99.81      & 97.13     & 91.93 \\
    \bottomrule[1.2pt]
    \end{tabular}}
    \end{threeparttable}
\end{table}

\subsection{Ablation Study}

We conducted ablation studies for both segmentation and classification tasks across four databases to validate the performance improvements provided by different modules, as shown in Table \ref{tbl9}. Specifically, we evaluated the impact of 1) removing the multi-resolution framework, i.e., only using the lowest resolution; 2) removing the continual learning module, i.e., training from scratch; 3) removing both the continual learning strategy and the multi-resolution framework; and 4) removing the decoder and the continual learning strategy, i.e., performing decoding at the lowest resolution. The experimental results confirm the applicability of our proposed framework for ECG and the benefits of our continual learning strategy for cross-domain incremental learning in ECG.

\section{Conclusion and Future Work}
In this study, we introduce a multi-resolution architecture serving as the groundwork for ECG interpretation tasks, along with a strategy based on parameter isolation for continual learning. With these foundational elements, we designed several types of continual learning experiments including domain incremental learning (transitioning from segmentation to classification tasks), task incremental learning (progressing from basic to complex tasks), and class incremental learning (advancing from tasks with a few classes to those with multiple classes). Experimental results showed that our proposed ECG-CL methodology significantly outperformed the traditional training-from-scratch approach. This suggests a potential for training a comprehensive ECG network capable of handling unobserved data and unique cardiac abnormalities. Further experiments with single-lead data demonstrated the prospective utility of our AI algorithms for smart wearable devices, enabling more thorough daily monitoring. Going forward, our goals include developing a model that supports knowledge backward transfer to better enhance prior tasks. We also aim to construct a task-free continual learning network that seamlessly integrates information from newly arriving data, thereby bolstering segmentation robustness and expanding classification categories.

\bibliographystyle{IEEEtran}
\bibliography{ref}

\end{document}